\documentclass[british,english,showpacs,preprintnumbers,amssymb,aps,notitlepage]{revtex4-1}
\usepackage{lmodern}

\usepackage[T1]{fontenc}
\usepackage[latin9]{inputenc}
\usepackage{color}
\usepackage{array}
\usepackage{float}
\usepackage{multirow}
\usepackage{amsbsy}
\usepackage{graphicx}
\PassOptionsToPackage{normalem}{ulem}
\usepackage{ulem}

\makeatletter

\providecommand{\tabularnewline}{\\}

\usepackage{amsfonts,amsmath} 

\usepackage[svgnames]{xcolor}
\usepackage{tikz}
\usetikzlibrary{decorations.markings}
\usetikzlibrary{shapes.geometric}

\newcommand\scaledLW{0.5} 

\newcommand\scaledNodeSize{0.5} 

\newcommand\nodescale{1.25} 

\pgfdeclarelayer{edgelayer}
\pgfdeclarelayer{nodelayer}
\pgfsetlayers{edgelayer,nodelayer,main}

\tikzstyle{none}=[inner sep=0pt]

\tikzstyle{rn}=[circle,fill=Red,draw=Black,line width=0.8 pt]
\tikzstyle{gn}=[circle,fill=Lime,draw=Black,line width=0.8 pt]
\tikzstyle{yn}=[circle,fill=Yellow,draw=Black,line width=0.8 pt]
\tikzstyle{auxiliary_qubit}=[circle,fill=Red,draw=none,scale=\nodescale,inner sep=3.5 pt, outer sep=-1pt]
\tikzstyle{logical_qubit}=[circle,fill=Black,draw=none,scale=\nodescale,inner sep=3.5 pt, outer sep=-1pt]
\tikzstyle{emb_logical_qubit}=[circle,fill=Black,draw=none,scale=\nodescale,inner sep=3.5 pt, outer sep=-1pt]
\tikzstyle{emb_auxiliary_qubit}=[circle,fill=Red,draw=none,scale=\nodescale,inner sep=3.5 pt, outer sep=-1pt]
\tikzstyle{unused_qubit}=[circle,fill=Gray,draw=none,scale=\nodescale,inner sep=3.5 pt, outer sep=-1pt]
\tikzstyle{arrow_end}=[circle,fill=none,draw=none,scale=.1]

\tikzstyle{simple}=[-,draw=Black,line width=1.000]
\tikzstyle{added}=[-,draw=Black,line width=1.000]
\definecolor{tempcolor}{rgb}{.9,.9,.9}
\tikzstyle{unused}=[-,draw=tempcolor,line width=0.500]
\tikzstyle{unused_added}=[-,draw=tempcolor,draw opacity=1,line width=0.5]
\tikzstyle{embedding}=[-,draw=Black,line width=3.250]
\tikzstyle{embedding_aux}=[-,draw=Red,line width=3.250]
\tikzstyle{arrow}=[-,draw=Black,postaction={decorate},decoration={markings,mark=at position .5 with {\arrow{>}}},line width=2.000]

\usepackage[font=small,labelfont=bf,
   justification=justified,
   singlelinecheck=false,
   format=plain]{caption}

\usepackage{threeparttable}

\usepackage{caption}
\usepackage[breaklinks]{hyperref}
\hypersetup{
    colorlinks,
    linkcolor={red},
    citecolor={magenta},
    urlcolor={blue}
}

\@ifundefined{showcaptionsetup}{}{%
 \PassOptionsToPackage{caption=false}{subfig}}
\usepackage{subfig}
\makeatother

\usepackage{babel}
\begin{document}

\title{Embedding quadratization gadgets on Chimera and Pegasus graphs}

\author{Nike Dattani}
\email{n.dattani@cfa.harvard.edu}

\affiliation{Harvard-Smithsonian Center for Astrophysics}

\author{Nicholas Chancellor}
\email{nicholas.chancellor@durham.ac.uk}

\affiliation{Durham University, Joint Quantum Centre}
\begin{abstract}
We group all known quadratizations of cubic and quartic terms in
binary optimization problems into five and six unique graphs respectively.
We then perform a minor embedding of these graphs onto the well-known
Chimera graph, and the brand new \emph{Pegasus} graph. We conclude
with recommendations for which gadgets are best to use when aiming
to reduce the total number of qubits required to embed a problem. 
\end{abstract}
\maketitle
Discrete optimization problems are often naturally formulated in terms
of minimizing some polynomial of degree $>2$ \citep{Dattani2014j},
which is then `quadratized' \citep{Dattani2019} into a quadratic
function which can be solved using standard algorithms for universal
classical computers \citep{Boros2002}, using special-purpose classical
annealers \citep{MasanaoYamaoka2016}, or using quantum annealers
\citep{Johnson2011}. With more than 40 quadratization methods available
for binary optimization problems \citep{Dattani2019}, one should
choose the best quadratization for a given problem, and for a given
method for solving the quadratized problem. 

There are ways to quadratize functions of discrete variables without
adding any auxiliary variables \citep{Ishikawa2014,Tanburn2015a,Okada2015s,Dridi2017,Dattani2019},
but when those methods cannot be applied we introduce auxiliary variables.
The resulting quadratic functions (called `gadgets'), that with good
accuracy (or sometimes exact accuracy) simulate the original high-degree
functions, will have some connectivity between the binary variables
(or bits, or qubits, herein referred to for convenience only, as qubits)
which can be represented by a graph in which vertices represent qubits
and edges indicate when two different qubits appear together in a
quadratic term. Since this graph incorporates no information about
the linear terms, constant term, or the coefficients of the quadratic
terms, many different gadgets have the same graph, therefore in this
paper we will classify all known quadratization gadgets into categories
according to their corresponding graph (herein called their `gadget
graph'). 

\noindent \vspace{-3mm}
\begin{figure}[h]
\caption{\label{fig:chimeraAndPegasus}Graph connectivities for D-Wave's Chimera
and Pegasus graphs.}
\vspace{-3mm}

\subfloat[Chimera (single cell)]{\includegraphics[width=0.23\textwidth]{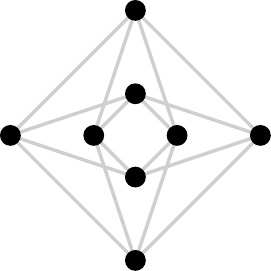}}~~~~\subfloat[Pegasus (single cell)]{

\includegraphics[width=0.23\textwidth]{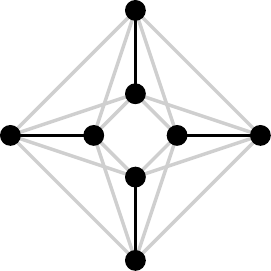}}~~~~\subfloat[Chimera (repeating)]{

\includegraphics[width=0.23\textwidth]{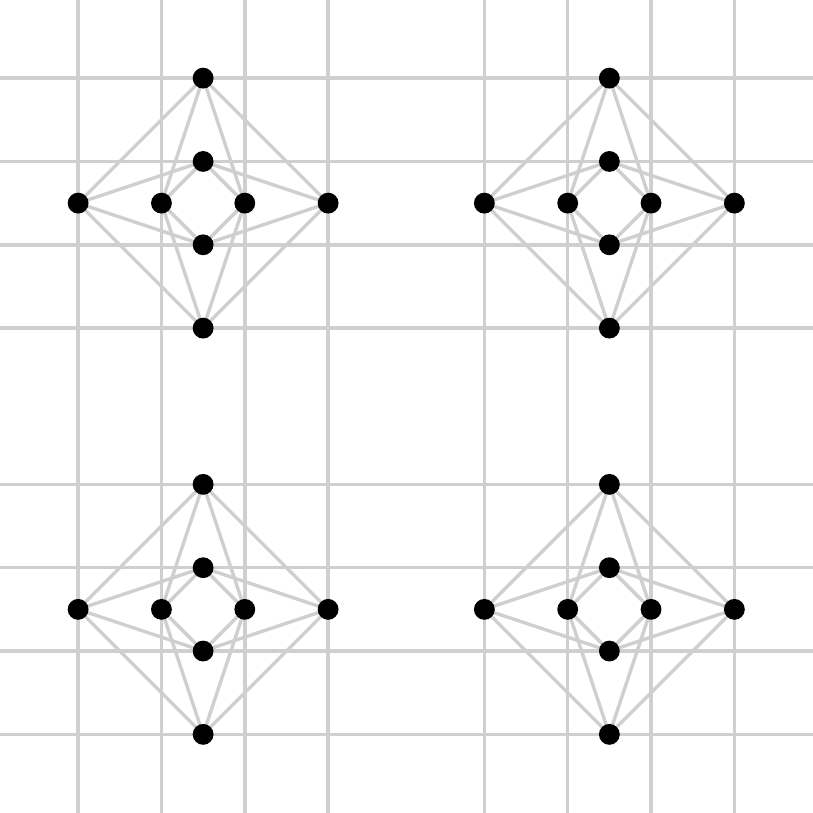}}~~~~\subfloat[Pegasus (repeating)]{

\includegraphics[width=0.23\textwidth]{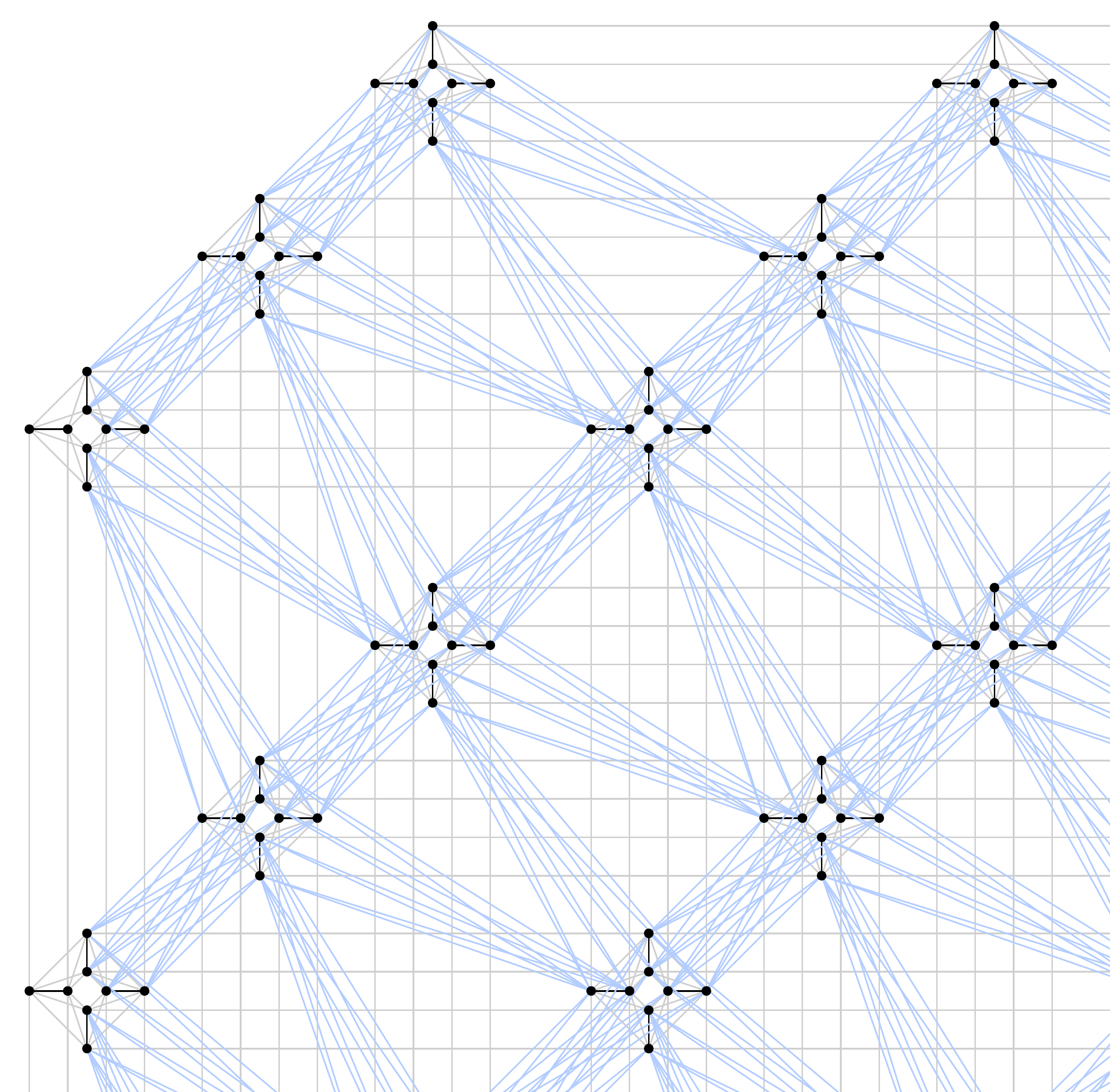}}
\end{figure}

\noindent \vspace{-5mm}

\noindent 

\noindent 
\begin{figure}[h]
{\small{}\caption{{\small{}\label{fig:allGraphGadgets}}\emph{\small{}Gadget graphs}{\small{}.
Graphs showing the connectivity between qubits in quadratization gadgets
for cubic to quadratic gadgets (top row), and quartic to quadratic
gadgets (bottom row). Red vertices represent auxiliary qubits and
black vertices represent logical qubits. Black edges denote the existence
of a quadratic term in the gadget, involving the two corresponding
qubits represented by vertices connected by the edge. }\emph{\small{}Linear
and constant terms in the gadgets are completely ignored here.}}
}{\small\par}

{\small{}\newcommand{\thisfigheight}{1.5 cm}}{\small\par}

\subfloat[\emph{\centering Propeller}]{

\includegraphics[width=0.12\textwidth]{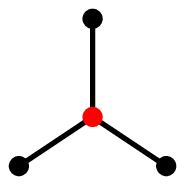}}~~\subfloat[\centering\emph{Coat Hanger}]{

\includegraphics[width=0.12\textwidth]{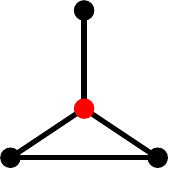}}~~\subfloat[\centering$K_{4}-e$]{

\includegraphics[width=0.12\textwidth]{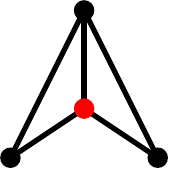}}~~\subfloat[\centering$K_{4}$]{

\includegraphics[width=0.12\textwidth]{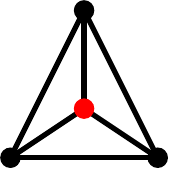}}~~\subfloat[\centering\emph{$K_{5}$ (2 aux)}]{

\includegraphics[width=0.12\textwidth]{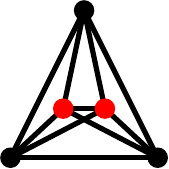}

}

\subfloat[\centering\emph{X}]{

\includegraphics[width=0.12\textwidth]{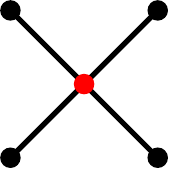}}~~~~\subfloat[\centering$K_{5}$\emph{ (1 aux)}]{

\includegraphics[width=0.12\textwidth]{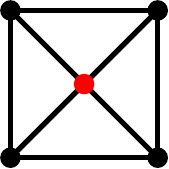}}~~~~\subfloat[\centering$K_{6}-4e$]{\includegraphics[width=0.12\textwidth]{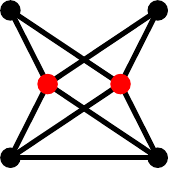}}~~~\subfloat[\centering\emph{$K_{6}-e$ }]{

\includegraphics[width=0.12\textwidth]{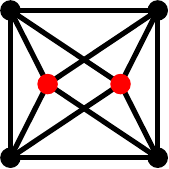}}~~~\subfloat[\centering\emph{$K_{6}$}]{ 

\includegraphics[width=0.12\textwidth]{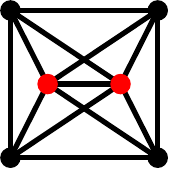}}~~~\subfloat[\centering\emph{Double }$K_{4}$]{

\includegraphics[width=0.12\textwidth]{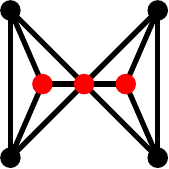}}
\end{figure}

\vspace{-3mm}

Gadget graphs for all known single cubic terms and for all known single
quartic terms are given in Fig. \ref{fig:allGraphGadgets}. Gadget
graphs tell us a lot about how costly the quadratic optimization problem
will be, and those with larger connectivity tend to yield more difficult
functions to optimize. Furthermore, some optimization methods only
work if their corresponding graph has a certain connectivity, two
examples of such connectivities being the ones in D-Wave's well-known
Chimera graph \citep{Neven2009}, and in their very recent \emph{Pegasus}
graph, both shown in Fig. \ref{fig:chimeraAndPegasus}.

\vspace{0mm}

\begin{table}[b]
\vspace{-2mm}

\caption{Grouping of all known quadratization gadgets for single \textbf{\emph{\uline{cubic}}}
terms, into categories corresponding to their gadget graphs. For each
unique gadget graph, the number of auxiliary qubits required for construction
of the gadget is listed, followed by the number of auxiliary qubits
required to minor-embed the gadget graph onto Chimera and Pegasus{\small{}.}}

\vspace{-2mm}

\rule{1\columnwidth}{0.5pt}

\begin{tabular*}{1\columnwidth}{@{\extracolsep{\fill}}cccccccccccc}
\hline 
\noalign{\vskip2mm}
\multirow{3}{*}{Gadget Graph} &  &  & \multirow{2}{*}{} & $N_{{\rm aux}}$ &  & $N_{{\rm aux}}$  & $N_{{\rm aux}}$  &  &  & $N_{{\rm aux}}$  & $N_{{\rm aux}}$ \tabularnewline
 & Example Gadgets & Reference &  & \textbf{\scriptsize{}Quadratization} & \multirow{2}{*}{} & \textbf{\scriptsize{}Embedding} & \textbf{\scriptsize{}Total} &  &  & \textbf{\scriptsize{}Embedding} & \textbf{\scriptsize{}Total}\tabularnewline[2mm]
\cline{7-8} \cline{11-12} 
\noalign{\vskip2mm}
 & (see \citep{Dattani2019} for definitions) &  &  &  &  & \multicolumn{2}{c}{\textbf{\textcolor{black}{Chimera}}} &  &  & \multicolumn{2}{c}{\textbf{\textcolor{black}{Pegasus}}}\tabularnewline[2mm]
\hline 
\noalign{\vskip2mm}
\multirow{2}{*}{\textbf{\textcolor{blue}{Propeller}}} & NTR-KZFD & \citep[Pg.7]{Dattani2019}\citep{Kolmogorov2004,Freedman2005} &  & \multirow{2}{*}{1} &  & \multirow{2}{*}{0} & \multirow{2}{*}{\textbf{\textcolor{red}{1}}} &  &  & \multirow{2}{*}{0} & \multirow{2}{*}{\textbf{\textcolor{red}{1}}}\tabularnewline
 & NTR-ABCG-2 & \citep[Pg.9]{Dattani2019}\citep{Anthony2016} &  &  &  &  &  &  &  &  & \tabularnewline[2mm]
\hline 
\noalign{\vskip2mm}
\multirow{1}{*}{\textbf{\textcolor{blue}{Coat hanger}}} & PTR-GBP & \citep[Pg.23]{Dattani2019}\citep{Gallagher2011} &  & \multirow{1}{*}{1} &  & \multirow{1}{*}{1} & \multirow{1}{*}{\textbf{\textcolor{red}{2}}} &  &  & \multirow{1}{*}{0} & \multirow{1}{*}{\textbf{\textcolor{red}{1}}}\tabularnewline[2mm]
\hline 
\noalign{\vskip2mm}
\multirow{2}{*}{\textbf{\textcolor{blue}{$\boldsymbol{\textcolor{blue}{\ensuremath{\boldsymbol{K_{4}}}-e}}$}}} & NTR-GBP & \citep[Pg.10]{Dattani2019}\citep{Gallagher2011} & \multirow{2}{*}{} & \multirow{2}{*}{1} & \multirow{2}{*}{} & \multirow{2}{*}{1} & \multirow{2}{*}{\textbf{\textcolor{red}{2}}} & \multirow{2}{*}{} & \multirow{2}{*}{} & \multirow{2}{*}{0} & \multirow{2}{*}{\textbf{\textcolor{red}{1}}}\tabularnewline
 & NTR-ABCG & \citep[Pg.8]{Dattani2019}\citep{Anthony2014} &  &  &  &  &  &  &  &  & \tabularnewline[2mm]
\hline 
\noalign{\vskip2mm}
\multirow{4}{*}{\textbf{\textcolor{blue}{$\boldsymbol{\textcolor{blue}{\ensuremath{\boldsymbol{K_{4}}}}}$}}} & PTR-Ishikawa & \citep[Pg.14]{Dattani2019}\citep{Ishikawa2011} &  & \multirow{4}{*}{1} &  & \multirow{4}{*}{2} & \multirow{4}{*}{\textbf{\textcolor{red}{3}}} & \multirow{4}{*}{} & \multirow{1}{*}{} & \multirow{4}{*}{0} & \multirow{4}{*}{\textbf{\textcolor{red}{1}}}\tabularnewline
 & NTR-RBL-(3$\rightarrow2$) & \citep[Pg.11]{Dattani2019}\citep{Rocchetto2016} &  &  &  &  &  &  & \multirow{3}{*}{} &  & \tabularnewline
 & PTR-BCR-1,2,3,4 & \citep[Pg.15-18]{Dattani2019}\citep{Boros2018,Boros2018a} &  &  &  &  &  &  &  &  & \tabularnewline
 & PTR-KZ & \citep[Pg.21]{Dattani2019}\citep{Kolmogorov2004,Chancellor2017,Leib2016} &  &  &  &  &  &  &  &  & \tabularnewline[2mm]
\hline 
\noalign{\vskip2mm}
\textbf{\textcolor{blue}{\emph{$\boldsymbol{\boldsymbol{K_{5}}}$}}}\textbf{\textcolor{blue}{{}
(2 aux)}} & PTR-RBL & \citep[Pg.24]{Dattani2019}\citep{Rocchetto2016} &  & 2 &  & 3 & \textbf{\textcolor{red}{5}} &  &  & 1 & \textbf{\textcolor{red}{2}}\tabularnewline[2mm]
\hline 
\end{tabular*}

\rule{1\columnwidth}{0.5pt}
\end{table}

\begin{table}[b]
\caption{{\small{}Grouping of all known quadratization gadgets for single }\textbf{\emph{\small{}\uline{quartic}}}{\small{}
terms, into categories corresponding to their gadget graphs. For each
unique gadget graph, the number of auxiliary qubits required for construction
of the gadget is listed, followed by the number of auxiliary qubits
required to minor-embed the gadget graph onto Chimera and Pegasus.}}

\rule{1\columnwidth}{0.5pt}

\begin{tabular*}{1\columnwidth}{@{\extracolsep{\fill}}cccccccccccc}
\hline 
\noalign{\vskip2mm}
\multirow{3}{*}{Gadget Graph} &  &  & \multirow{2}{*}{} & $N_{{\rm aux}}$ &  & $N_{{\rm aux}}$  & $N_{{\rm aux}}$  &  &  & $N_{{\rm aux}}$  & $N_{{\rm aux}}$ \tabularnewline
 & Example Gadgets & Reference &  & \textbf{\scriptsize{}Quadratization} & \multirow{2}{*}{} & \textbf{\scriptsize{}Embedding} & \textbf{\scriptsize{}Total} &  &  & \textbf{\scriptsize{}Embedding} & \textbf{\scriptsize{}Total}\tabularnewline[2mm]
\cline{7-8} \cline{11-12} 
\noalign{\vskip2mm}
 & (see \citep{Dattani2019} for definitions) &  &  &  &  & \multicolumn{2}{c}{\textbf{\textcolor{black}{Chimera}}} &  &  & \multicolumn{2}{c}{\textbf{\textcolor{black}{Pegasus}}}\tabularnewline[2mm]
\hline 
\noalign{\vskip2mm}
\multirow{2}{*}{\textbf{\textcolor{blue}{\emph{X}}}} & NTR-KZFD & \citep[Pg.7]{Dattani2019}\citep{Kolmogorov2004,Freedman2005} &  & \multirow{2}{*}{1} &  & \multirow{2}{*}{0} & \multirow{2}{*}{\textbf{\textcolor{red}{1}}} &  &  & \multirow{2}{*}{0} & \multirow{2}{*}{\textbf{\textcolor{red}{1}}}\tabularnewline
 & NTR-ABCG-2 & \citep[Pg.9]{Dattani2019}\citep{Anthony2016} &  &  &  &  &  &  &  &  & \tabularnewline[2mm]
\hline 
\noalign{\vskip2mm}
\multirow{3}{*}{\textbf{\textcolor{blue}{\emph{$\boldsymbol{K_{5}}$ }}}\textbf{\textcolor{blue}{(1
aux)}}} & PTR-BCR-2 & \citep[Pg.16]{Dattani2019}\citep{Boros2018,Boros2018a} &  & \multirow{3}{*}{1} &  & \multirow{3}{*}{3} & \multirow{3}{*}{\textbf{\textcolor{red}{4}}} &  &  & \multirow{3}{*}{1} & \multirow{3}{*}{\textbf{\textcolor{red}{2}}}\tabularnewline
 & PTR-BCR-4 & \citep[Pg.18]{Dattani2019}\citep{Boros2018,Boros2018a} &  &  &  &  &  &  &  &  & \tabularnewline
 & NTR-LHZ & \citep[Pg.12]{Dattani2019}\citep{Lechner2015a} &  &  &  &  &  &  &  &  & \tabularnewline[2mm]
\hline 
\noalign{\vskip2mm}
\textbf{\textcolor{blue}{$\boldsymbol{\boldsymbol{K_{6}}-4e}$}} & PTR-BG & \citep[Pg.13]{Dattani2019}\citep{Boros2014} &  & 2 &  & 1 & \textbf{\textcolor{red}{3}} &  &  & 0 & \textbf{\textcolor{red}{2}}\tabularnewline[2mm]
\hline 
\noalign{\vskip2mm}
\textbf{\textcolor{blue}{$\boldsymbol{\boldsymbol{K_{6}}-e}$}} & PTR-Ishikawa & \citep[Pg.14]{Dattani2019}\citep{Ishikawa2011} &  & 2 &  & 5 & \textbf{\textcolor{red}{7}} &  &  & 2 & \textbf{\textcolor{red}{4}}\tabularnewline[2mm]
\hline 
\noalign{\vskip2mm}
\textbf{\textcolor{blue}{$\boldsymbol{\boldsymbol{K_{6}}}$}} & PTR-BCR-3 & \citep[Pg.17]{Dattani2019}\citep{Boros2018,Boros2018a} &  & 2 &  & 8 & \textbf{\textcolor{red}{10}} &  &  & 2 & \textbf{\textcolor{red}{4}}\tabularnewline[2mm]
\hline 
\noalign{\vskip2mm}
\textbf{\textcolor{blue}{Double }}\textbf{\textcolor{blue}{\emph{$\boldsymbol{\boldsymbol{K_{4}}}$}}} & PTR-LZL & \citep[Pg.26]{Dattani2019}\citep{Chancellor2017} &  & 3 &  & 5 & \textbf{\textcolor{red}{8}} &  &  & 1 & \textbf{\textcolor{red}{4}}\tabularnewline[2mm]
\hline 
\end{tabular*}

\rule{1\columnwidth}{0.5pt}
\end{table}

Any graph, can be mapped onto the Chimera or Pegasus graphs by minor-embedding
\citep{Choi2008,Choi2011}, where the Chimera graph or the Pegasus
graph is a graph minor of the graph representing the problem that
needs to be optimized. This often means that one binary variable in
the quadratic optimization problem needs to be represented by two
(or more) qubits (called a `chain') instead of one, making the number
of physical qubits needed to solve the original problem larger than
before, and sometimes impossibly large to solve. For example if we
wanted to factor the smallest RSA number which has not yet been factored
on a classical computer (RSA-230) by minimization of a corresponding
binary optimization problem, we would have to minimize a quartic function
of 6594 variables which can be turned into a quadratic problem involving
148~776 variables \citep{Li2017}. Say we have a quantum annealer
with 149~000 qubits: This means that only 224 out of the 142~182
required gadget graphs can afford an auxiliary qubit for minor-embedding
into the hardware's native connectivity graph. We therefore want \emph{efficient
gadgets} which can be minor-embedded into our hardware with \emph{zero}
or \emph{very few} extra qubits. 

In this paper we have provided minor-embeddings for all gadget graphs
in Fig. \ref{fig:allGraphGadgets}, for both Chimera and Pegasus.
We note that \textbf{\emph{all}} cubic to quadratic gadgets involving
one auxiliary qubit can be embedded onto Pegasus without any further
auxiliary qubits for the embedding, because Pegasus contains the $K_{4}$
graph, which means any possible connections between the three logical
qubits and the one auxiliary qubit are already contained in Pegasus.
Since Chimera does not contain $K_{4}$, only negative cubic terms
are so far known to be quadratizable with gadgets that embed directly
onto Chimera without any extra qubits for the embedding. Furthermore,
\emph{all} cubic and quartic gadgets are embeddable with a single
`cell' of Pegasus and with chains of length at most two, whereas with
Chimera cells, three of the quartic gadgets require multiple cells
and two require chains of length three.

\vspace{-3mm}

\section{Minor embeddings for cubic to quadratic gadgets}

\vspace{-4mm}

\subsection{Chimera graph}

\vspace{-5mm}

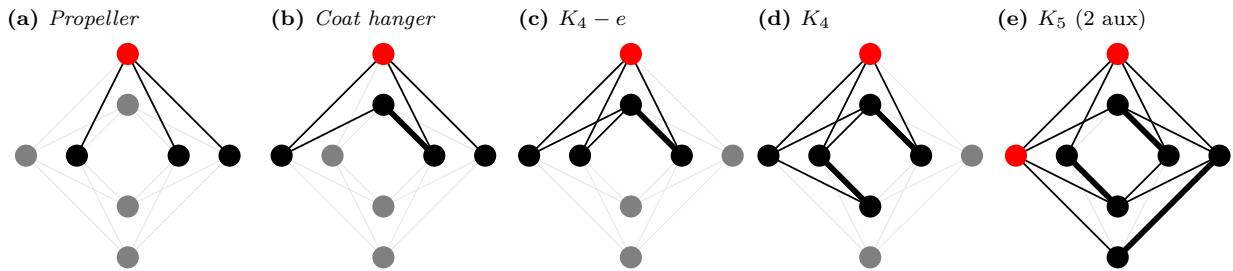
\begin{figure}[H]
\caption{{\small{}Minor embeddings of all }\textbf{\emph{\small{}\uline{cubic}}}\textbf{\emph{\small{}
}}{\small{}to quadratic gadgets for single terms, onto a `unit cell'
of a Chimera graph. Grey vertices and edges are not used. Thick edges
denote chains for minor embedding, in which two or more qubits (vertices)
represent one logical qubit (this is done when logical qubits need
to be connected to more qubits than the Chimera unit cell otherwise
allows).}}

\newcommand{\chimeraFigSize}{3cm}
\centering{}\subfloat[\emph{Propeller\label{fig:all-to-auxChimera}}]{

\resizebox{!}{\chimeraFigSize}{\begin{tikzpicture}
	\begin{pgfonlayer}{nodelayer}
		\node [style={auxiliary_qubit}] (0) at (0, 2) {};
		\node [style={unused_qubit}] (1) at (0, 1) {};
		\node [style={unused_qubit}] (2) at (0, -2) {};
		\node [style={logical_qubit}] (3) at (1, 0) {};
		\node [style={logical_qubit}] (4) at (-1, 0) {};
		\node [style={unused_qubit}] (5) at (-2, 0) {};
		\node [style={unused_qubit}] (6) at (0, -1) {};
		\node [style={logical_qubit}] (7) at (2, 0) {};
	\end{pgfonlayer}
	\begin{pgfonlayer}{edgelayer}
		\draw [style=unused] (1) to (3);
		\draw [style=unused] (0) to (5);
		\draw [style=unused] (5) to (2);
		\draw [style=unused] (5) to (1);
		\draw [style=unused] (6) to (3);
		\draw [style=unused] (6) to (4);
		\draw [style=unused] (6) to (5);
		\draw [style=unused] (6) to (7);
		\draw [style=unused] (7) to (2);
		\draw [style=unused] (1) to (7);
		\draw [style=unused] (3) to (2);
		\draw [style=unused] (1) to (4);
		\draw [style=unused] (4) to (2);
		\draw [style=simple] (0) to (7);
		\draw [style=simple] (0) to (3);
		\draw [style=simple] (4) to (0);
	\end{pgfonlayer}
\end{tikzpicture}}}~~~\subfloat[\emph{Coat hanger}]{\resizebox{!}{\chimeraFigSize}{\begin{tikzpicture}
	\begin{pgfonlayer}{nodelayer}
		\node [style={logical_qubit}] (0) at (-2, 0) {};
		\node [style={unused_qubit}] (1) at (-1, 0) {};
		\node [style={logical_qubit}] (2) at (2, 0) {};
		\node [style={emb_logical_qubit}] (3) at (0, 1) {};
		\node [style={unused_qubit}] (4) at (0, -1) {};
		\node [style={unused_qubit}] (5) at (0, -2) {};
		\node [style={emb_logical_qubit}] (6) at (1, 0) {};
		\node [style={auxiliary_qubit}] (7) at (0, 2) {};
	\end{pgfonlayer}
	\begin{pgfonlayer}{edgelayer}
		\draw [style=unused] (1) to (3);
		\draw [style=unused] (3) to (2);
		\draw [style=unused] (1) to (4);
		\draw [style=unused] (4) to (2);
		\draw [style=unused] (4) to (0);
		\draw [style=unused] (0) to (5);
		\draw [style=unused] (5) to (2);
		\draw [style=unused] (5) to (1);
		\draw [style=embedding] (6) to (3);
		\draw [style=unused] (6) to (4);
		\draw [style=unused] (6) to (5);
		\draw [style=simple] (0) to (7);
		\draw [style=simple] (6) to (7);
		\draw [style=simple] (7) to (2);
		\draw [style=unused] (1) to (7);
		\draw [style=simple] (0) to (3);
	\end{pgfonlayer}
\end{tikzpicture}}}~~\subfloat[$K_{4}-e$]{\resizebox{!}{\chimeraFigSize}{\begin{tikzpicture}
	\begin{pgfonlayer}{nodelayer}
		\node [style={logical_qubit}] (0) at (-2, -0) {};
		\node [style={logical_qubit}] (1) at (-1, -0) {};
		\node [style=unused_qubit] (2) at (2, -0) {};
		\node [style={emb_logical_qubit}] (3) at (0, 1) {};
		\node [style=unused_qubit] (4) at (0, -1) {};
		\node [style=unused_qubit] (5) at (0, -2) {};
		\node [style={emb_logical_qubit}] (6) at (1, -0) {};
		\node [style={auxiliary_qubit}] (7) at (0, 2) {};
	\end{pgfonlayer}
	\begin{pgfonlayer}{edgelayer}
		\draw [style=simple] (1) to (3);
		\draw [style=simple] (0) to (3);
		\draw [style=unused] (3) to (2);
		\draw [style=unused] (1) to (4);
		\draw [style=unused] (4) to (2);
		\draw [style=unused] (4) to (0);
		\draw [style=unused] (0) to (5);
		\draw [style=unused] (5) to (2);
		\draw [style=unused] (5) to (1);
		\draw [style={embedding}] (6) to (3);
		\draw [style=unused] (6) to (4);
		\draw [style=unused] (6) to (5);
		\draw [style=simple] (0) to (7);
		\draw [style=simple] (6) to (7);
		\draw [style=unused] (7) to (2);
		\draw [style=simple] (1) to (7);
	\end{pgfonlayer}
\end{tikzpicture}}}~\subfloat[$K_{4}$]{\resizebox{!}{\chimeraFigSize}{\begin{tikzpicture}
	\begin{pgfonlayer}{nodelayer}
		\node [style={logical_qubit}] (0) at (-2, -0) {};
		\node [style={emb_logical_qubit}] (1) at (-1, -0) {};
		\node [style=unused_qubit] (2) at (2, -0) {};
		\node [style={emb_logical_qubit}] (3) at (0, 1) {};
		\node [style={emb_logical_qubit}] (4) at (0, -1) {};
		\node [style=unused_qubit] (5) at (0, -2) {};
		\node [style={emb_logical_qubit}] (6) at (1, -0) {};
		\node [style={auxiliary_qubit}] (7) at (0, 2) {};
	\end{pgfonlayer}
	\begin{pgfonlayer}{edgelayer}
		\draw [style=unused] (0) to (5);
		\draw [style=unused] (5) to (2);
		\draw [style=unused] (5) to (1);
		\draw [style=unused] (6) to (4);
		\draw [style=unused] (6) to (5);
		\draw [style=unused] (4) to (2);
		\draw [style=unused] (7) to (2);
		\draw [style=unused] (3) to (2);
		\draw [style=simple] (1) to (3);
		\draw [style=simple] (0) to (3);
		
		\draw [style={embedding}] (1) to (4);
		
		\draw [style=simple] (4) to (0);
		
		\draw [style={embedding}] (6) to (3);
		
		\draw [style=simple] (0) to (7);
		\draw [style=simple] (6) to (7);
		
		\draw [style=simple] (1) to (7);
	\end{pgfonlayer}
\end{tikzpicture}}}~\subfloat[\emph{$K_{5}$} (2 aux)]{

\resizebox{!}{\chimeraFigSize}{\begin{tikzpicture}
	\begin{pgfonlayer}{nodelayer}
		\node [style={auxiliary_qubit}] (0) at (-2, 0) {};
		\node [style={emb_logical_qubit}] (1) at (-1, 0) {};
		\node [style={emb_logical_qubit}] (2) at (2, 0) {};
		\node [style={auxiliary_qubit}] (3) at (0, 2) {};
		\node [style={emb_logical_qubit}] (4) at (0, 1) {};
		\node [style={emb_logical_qubit}] (5) at (0, -1) {};
		\node [style={emb_logical_qubit}] (6) at (0, -2) {};
		\node [style={emb_logical_qubit}] (7) at (1, 0) {};
	\end{pgfonlayer}
	\begin{pgfonlayer}{edgelayer}
		\draw [style=simple] (3) to (2);
		\draw [style=unused] (1) to (4);
		\draw [style=simple] (0) to (4);
		\draw [style=simple] (4) to (2);
		\draw [style=embedding] (1) to (5);
		\draw [style=simple] (0) to (6);
		\draw [style=embedding] (6) to (2);
		\draw [style=unused] (6) to (1);
		\draw [style=simple] (0) to (3);
		\draw [style=simple] (1) to (3);
		\draw [style=embedding] (7) to (4);
		\draw [style=simple] (7) to (5);
		\draw [style=unused] (7) to (6);
		\draw [style=simple] (7) to (3);
		\draw [style=simple] (5) to (2);
		\draw [style=simple] (0) to (5);
	\end{pgfonlayer}
\end{tikzpicture}}}
\end{figure}

\vspace{-6mm}

\subsection{Pegasus graph}

\vspace{-6mm}

\begin{figure}[H]
\caption{Minor embeddings of all \textbf{\emph{\uline{cubic}}}\textbf{\emph{
}}to quadratic gadgets for single terms, onto a Pegasus `cell'. Grey
vertices and edges are not used.}

\newcommand{\chimeraFigSize}{3cm}
\centering{}\subfloat[\emph{Propeller}]{\resizebox{!}{\chimeraFigSize}{\begin{tikzpicture}
	\begin{pgfonlayer}{nodelayer}
		\node [style={auxiliary_qubit}] (0) at (0, 2) {};
		\node [style={unused_qubit}] (1) at (0, 1) {};
		\node [style={unused_qubit}] (2) at (0, -2) {};
		\node [style={logical_qubit}] (3) at (1, 0) {};
		\node [style={logical_qubit}] (4) at (-1, 0) {};
		\node [style={unused_qubit}] (5) at (-2, 0) {};
		\node [style={unused_qubit}] (6) at (0, -1) {};
		\node [style={logical_qubit}] (7) at (2, 0) {};
	\end{pgfonlayer}
	\begin{pgfonlayer}{edgelayer}
		\draw [style=unused] (6) to (7);
		\draw [style=unused] (7) to (2);
		\draw [style=unused] (1) to (7);
		\draw [style=unused] (1) to (3);
		\draw [style=unused] (0) to (5);
		\draw [style=unused] (5) to (2);
		\draw [style=unused] (5) to (1);
		\draw [style=unused] (6) to (3);
		\draw [style=unused] (6) to (4);
		\draw [style=unused] (6) to (5);
		\draw [style=unused] (3) to (2);
		\draw [style=unused] (1) to (4);
		\draw [style=unused] (4) to (2);
		\draw [style=simple] (0) to (3);
		\draw [style=simple] (4) to (0);
		\draw [style=simple] (0) to (7);
		\draw [style={unused_added}] (0) to (1);
		\draw [style={unused_added}] (7) to (3);
		\draw [style={unused_added}] (5) to (4);
		\draw [style={unused_added}] (6) to (2);
	\end{pgfonlayer}
\end{tikzpicture}}}~~\subfloat[\emph{Coat hanger}]{\resizebox{!}{\chimeraFigSize}{\begin{tikzpicture}
	\begin{pgfonlayer}{nodelayer}
		\node [style={logical_qubit}] (0) at (2, 0) {};
		\node [style={logical_qubit}] (1) at (1, 0) {};
		\node [style={unused_qubit}] (2) at (-2, 0) {};
		\node [style={unused_qubit}] (3) at (0, -2) {};
		\node [style={unused_qubit}] (4) at (0, -1) {};
		\node [style={logical_qubit}] (5) at (0, 1) {};
		\node [style={auxiliary_qubit}] (6) at (0, 2) {};
		\node [style={unused_qubit}] (7) at (-1, 0) {};
	\end{pgfonlayer}
	\begin{pgfonlayer}{edgelayer}
		\draw [style=unused] (3) to (2);
		\draw [style=unused] (1) to (4);
		\draw [style=unused] (0) to (4);
		\draw [style=unused] (4) to (2);
		\draw [style={unused_added}] (0) to (1);
		\draw [style=unused] (1) to (5);
		\draw [style=unused] (5) to (2);
		\draw [style=simple] (5) to (0);
		\draw [style=simple] (0) to (6);
		\draw [style=unused] (6) to (2);
		\draw [style=simple] (6) to (1);
		\draw [style=added] (5) to (6);
		\draw [style=unused] (0) to (3);
		\draw [style=unused] (1) to (3);
		\draw [style=unused] (7) to (4);
		\draw [style=unused] (7) to (5);
		\draw [style=unused] (7) to (6);
		\draw [style=unused] (7) to (3);
		\draw [style={unused_added}] (7) to (2);
		\draw [style={unused_added}] (3) to (4);
	\end{pgfonlayer}
\end{tikzpicture}}}~\subfloat[\emph{$K_{4}-e$}]{	\resizebox{!}{\chimeraFigSize}{\begin{tikzpicture}
	\begin{pgfonlayer}{nodelayer}
		\node [style={auxiliary_qubit}] (0) at (0, 2) {};
		\node [style={logical_qubit}] (1) at (0, 1) {};
		\node [style={unused_qubit}] (2) at (0, -2) {};
		\node [style={unused_qubit}] (3) at (-2, 0) {};
		\node [style={unused_qubit}] (4) at (-1, 0) {};
		\node [style={logical_qubit}] (5) at (1, 0) {};
		\node [style={logical_qubit}] (6) at (2, 0) {};
		\node [style={unused_qubit}] (7) at (0, -1) {};
	\end{pgfonlayer}
	\begin{pgfonlayer}{edgelayer}
		\draw [style=unused] (3) to (2);
		\draw [style=unused] (1) to (4);
		\draw [style=unused] (0) to (4);
		\draw [style=unused] (4) to (2);
		\draw [style=added] (0) to (1);
		\draw [style=unused] (1) to (5);
		\draw [style=unused] (5) to (2);
		\draw [style=simple] (5) to (0);
		\draw [style=simple] (0) to (6);
		\draw [style=unused] (6) to (2);
		\draw [style=simple] (6) to (1);
		\draw [style=added] (5) to (6);
		\draw [style=unused] (0) to (3);
		\draw [style=unused] (1) to (3);
		\draw [style=unused] (7) to (4);
		\draw [style=unused] (7) to (5);
		\draw [style=unused] (7) to (6);
		\draw [style=unused] (7) to (3);
		\draw [style={unused_added}] (7) to (2);
		\draw [style={unused_added}] (3) to (4);
	\end{pgfonlayer}
\end{tikzpicture}}}~~\subfloat[\emph{$K_{4}$ }]{\resizebox{!}{\chimeraFigSize}{\begin{tikzpicture}
	\begin{pgfonlayer}{nodelayer}
		\node [style={logical_qubit}] (0) at (2, 0) {};
		\node [style={logical_qubit}] (1) at (1, 0) {};
		\node [style={unused_qubit}] (2) at (-2, 0) {};
		\node [style={unused_qubit}] (3) at (0, -2) {};
		\node [style={unused_qubit}] (4) at (0, -1) {};
		\node [style={logical_qubit}] (5) at (0, 1) {};
		\node [style={auxiliary_qubit}] (6) at (0, 2) {};
		\node [style={unused_qubit}] (7) at (-1, 0) {};
	\end{pgfonlayer}
	\begin{pgfonlayer}{edgelayer}
		\draw [style=unused] (3) to (2);
		\draw [style=unused] (1) to (4);
		\draw [style=unused] (0) to (4);
		\draw [style=unused] (4) to (2);
		\draw [style=added] (0) to (1);
		\draw [style=simple] (1) to (5);
		\draw [style=unused] (5) to (2);
		\draw [style=simple] (5) to (0);
		\draw [style=simple] (0) to (6);
		\draw [style=unused] (6) to (2);
		\draw [style=simple] (6) to (1);
		\draw [style=added] (5) to (6);
		\draw [style=unused] (0) to (3);
		\draw [style=unused] (1) to (3);
		\draw [style=unused] (7) to (4);
		\draw [style=unused] (7) to (5);
		\draw [style=unused] (7) to (6);
		\draw [style=unused] (7) to (3);
		\draw [style={unused_added}] (7) to (2);
		\draw [style={unused_added}] (3) to (4);
	\end{pgfonlayer}
\end{tikzpicture}}}~~\subfloat[\emph{$K_{5}$ (2 aux)}]{\resizebox{!}{\chimeraFigSize}{\begin{tikzpicture}
	\begin{pgfonlayer}{nodelayer}
		\node [style={auxiliary_qubit}] (0) at (-2, 0) {};
		\node [style={logical_qubit}] (1) at (-1, 0) {};
		\node [style={unused_qubit}] (2) at (2, 0) {};
		\node [style={auxiliary_qubit}] (3) at (0, 2) {};
		\node [style={logical_qubit}] (4) at (0, 1) {};
		\node [style={emb_logical_qubit}] (5) at (0, -1) {};
		\node [style={unused_qubit}] (6) at (0, -2) {};
		\node [style={emb_logical_qubit}] (7) at (1, 0) {};
	\end{pgfonlayer}
	\begin{pgfonlayer}{edgelayer}
		\draw [style=unused] (3) to (2);
		\draw [style=unused] (4) to (2);
		\draw [style=unused] (0) to (6);
		\draw [style=unused] (6) to (2);
		\draw [style=unused] (6) to (1);
		\draw [style=unused] (5) to (2);
		\draw [style=unused] (7) to (6);
		\draw [style=simple] (1) to (4);
		\draw [style=simple] (0) to (4);
		\draw [style=added] (0) to (1);
		\draw [style=simple] (1) to (5);
		\draw [style=simple] (5) to (0);
		\draw [style=added] (3) to (4);
		\draw [style={unused_added}] (5) to (6);
		\draw [style=simple] (0) to (3);
		\draw [style=simple] (1) to (3);
		\draw [style=simple] (7) to (4);
		\draw [style=embedding] (7) to (5);
		\draw [style=simple] (7) to (3);
		\draw [style={unused_added}] (7) to (2);
	\end{pgfonlayer}
\end{tikzpicture}}}
\end{figure}
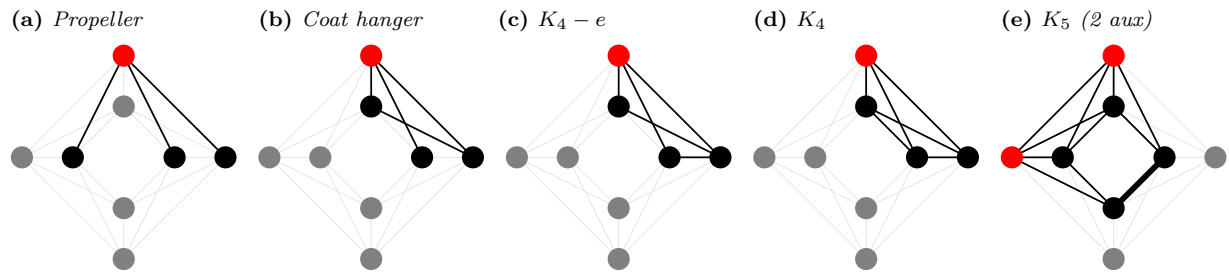

\vspace{-3mm}

\section{Minor embeddings for quartic to quadratic gadgets}

\vspace{-1mm}

\subsection{Chimera graph}

\vspace{-5mm}

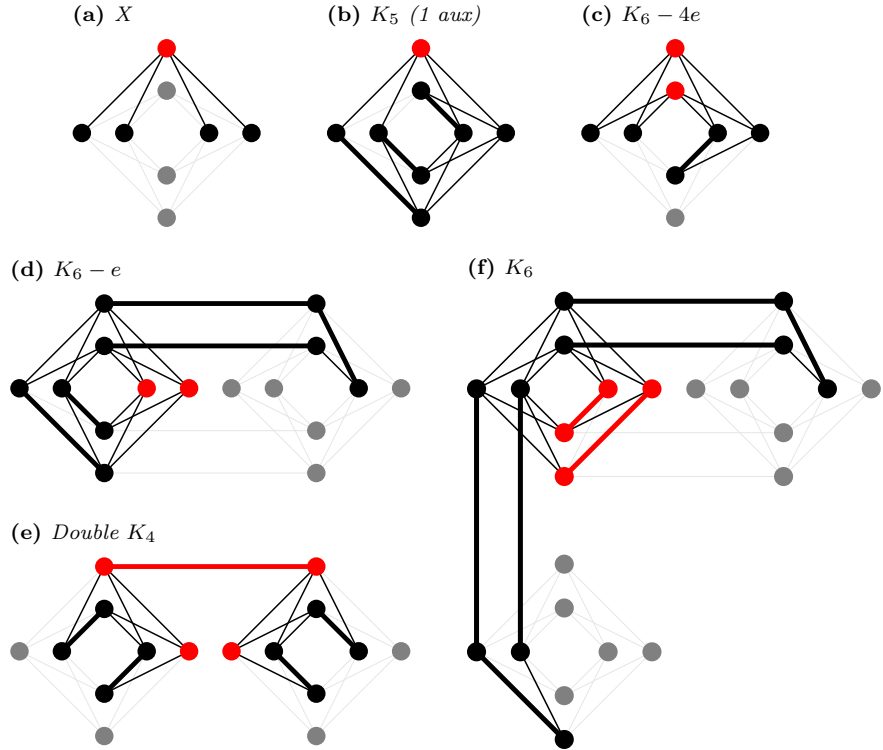
\begin{figure}[H]
\caption{{\small{}Minor embeddings of all }\textbf{\emph{\small{}\uline{quartic}}}{\small{}
to quadratic gadgets for single terms, onto a `unit cell' of a Chimera
graph. Grey vertices and edges are not used. Thick edges denote chains
for minor embedding, in which two or more qubits (vertices) represent
one logical qubit (this is done when logical qubits need to be connected
to more qubits than the Chimera unit cell otherwise allows).}}

\newcommand{\chimeraFigSize}{2.5cm}

\noindent\begin{minipage}[t]{1\textwidth}%
\begin{center}
\par\end{center}
\begin{center}
\hspace{-0.9 cm}\subfloat[\emph{X\label{fig:all-to-auxChimera-1}}]{\resizebox{!}{\chimeraFigSize}{\begin{tikzpicture}
	\begin{pgfonlayer}{nodelayer}
		\node [style=logical_qubit] (0) at (-2, -0) {};
		\node [style=logical_qubit] (1) at (-1, -0) {};
		\node [style=logical_qubit] (2) at (1, -0) {};
		\node [style=logical_qubit] (3) at (2, -0) {};
		\node [style=auxiliary_qubit] (4) at (0, 2) {};
		\node [style=unused_qubit] (5) at (0, 1) {};
		\node [style=unused_qubit] (6) at (0, -1) {};
		\node [style=unused_qubit] (7) at (0, -2) {};
	\end{pgfonlayer}
	\begin{pgfonlayer}{edgelayer}
		\draw [style=unused] (1) to (5);
		\draw [style=unused] (2) to (5);
		\draw [style=unused] (0) to (5);
		\draw [style=unused] (5) to (3);
		\draw [style=unused] (1) to (6);
		\draw [style=unused] (6) to (2);
		\draw [style=unused] (6) to (3);
		\draw [style=unused] (6) to (0);
		\draw [style=unused] (0) to (7);
		\draw [style=unused] (7) to (3);
		\draw [style=unused] (7) to (2);
		\draw [style=unused] (7) to (1);
		\draw [style=simple] (0) to (4);
		\draw [style=simple] (1) to (4);
		\draw [style=simple] (2) to (4);
		\draw [style=simple] (4) to (3);
		
	\end{pgfonlayer}
\end{tikzpicture}
}}~~~~~~~\subfloat[\textcolor{black}{$K_{5}$}\textcolor{black}{\emph{ (1 aux)}}]{\resizebox{!}{\chimeraFigSize}{\begin{tikzpicture}
	\begin{pgfonlayer}{nodelayer}
		\node [style=emb_logical_qubit] (0) at (-2, -0) {};
		\node [style=emb_logical_qubit] (1) at (-1, -0) {};
		\node [style=logical_qubit] (2) at (2, -0) {};
		\node [style=auxiliary_qubit] (3) at (0, 2) {};
		\node [style=emb_logical_qubit] (4) at (0, 1) {};
		\node [style=emb_logical_qubit] (5) at (0, -1) {};
		\node [style=emb_logical_qubit] (6) at (0, -2) {};
		\node [style=emb_logical_qubit] (7) at (1, -0) {};
	\end{pgfonlayer}
	\begin{pgfonlayer}{edgelayer}
		\draw [style=simple] (3) to (2);
		\draw [style=unused] (1) to (4);
		\draw [style=unused] (0) to (4);
		\draw [style=simple] (4) to (2);
		\draw [style={embedding}] (1) to (5);
		\draw [style=simple] (5) to (2);
		\draw [style=unused] (5) to (0);
		\draw [style={embedding}] (0) to (6);
		\draw [style=simple] (6) to (2);
		\draw [style=simple] (6) to (1);
		\draw [style=simple] (0) to (3);
		\draw [style=simple] (1) to (3);
		\draw [style={embedding}] (7) to (4);
		\draw [style=simple] (7) to (5);
		\draw [style=simple] (7) to (6);
		\draw [style=simple] (7) to (3);
	\end{pgfonlayer}
\end{tikzpicture}
}}~~~~~~~\subfloat[\emph{$K_{6}-4e$}]{\resizebox{!}{\chimeraFigSize}{\begin{tikzpicture}
	\begin{pgfonlayer}{nodelayer}
		\node [style=logical_qubit] (0) at (-2, -0) {};
		\node [style=logical_qubit] (1) at (-1, -0) {};
		\node [style=logical_qubit] (2) at (2, -0) {};
		\node [style=auxiliary_qubit] (3) at (0, 2) {};
		\node [style=auxiliary_qubit] (4) at (0, 1) {};
		\node [style=emb_logical_qubit] (5) at (0, -1) {};
		\node [style=unused_qubit] (6) at (0, -2) {};
		\node [style=emb_logical_qubit] (7) at (1, -0) {};
	\end{pgfonlayer}
	\begin{pgfonlayer}{edgelayer}
		\draw [style=unused] (0) to (6);
		\draw [style=unused] (6) to (2);
		\draw [style=unused] (6) to (1);
		\draw [style=unused] (1) to (5);
		\draw [style=unused] (7) to (6);
		\draw [style=unused] (5) to (0);
		\draw [style=simple] (3) to (2);
		\draw [style=simple] (1) to (4);
		\draw [style=simple] (0) to (4);
		\draw [style=simple] (4) to (2);
		
		\draw [style=simple] (5) to (2);

		\draw [style=simple] (0) to (3);
		\draw [style=simple] (1) to (3);
		\draw [style=simple] (7) to (4);
		\draw [style={embedding}] (7) to (5);
		
		\draw [style=simple] (7) to (3);
	\end{pgfonlayer}
\end{tikzpicture}
}}
\par\end{center}%
\end{minipage}

\hfill{}%
\begin{minipage}[t]{0.3\textwidth}%
\vspace{-0mm}

\begin{center}
\subfloat[\emph{$K_{6}-e$}]{\resizebox{!}{\chimeraFigSize}{\begin{tikzpicture}
	\begin{pgfonlayer}{nodelayer}
		\node [style={emb_logical_qubit}] (0) at (0, -2) {};
		\node [style={emb_logical_qubit}] (1) at (0, -1) {};
		\node [style={emb_logical_qubit}] (2) at (0, 2) {};
		\node [style={auxiliary_qubit}] (3) at (2, 0) {};
		\node [style={auxiliary_qubit}] (4) at (1, 0) {};
		\node [style={emb_logical_qubit}] (5) at (-1, 0) {};
		\node [style={emb_logical_qubit}] (6) at (-2, 0) {};
		\node [style={emb_logical_qubit}] (7) at (0, 1) {};
		\node [style={unused_qubit}] (8) at (5, -2) {};
		\node [style={unused_qubit}] (9) at (5, -1) {};
		\node [style={emb_logical_qubit}] (10) at (5, 2) {};
		\node [style={unused_qubit}] (11) at (7, 0) {};
		\node [style={emb_logical_qubit}] (12) at (6, 0) {};
		\node [style={unused_qubit}] (13) at (4, 0) {};
		\node [style={unused_qubit}] (14) at (3, 0) {};
		\node [style={emb_logical_qubit}] (15) at (5, 1) {};
	\end{pgfonlayer}
	\begin{pgfonlayer}{edgelayer}
		\draw [style=unused] (15) to (13);
		\draw [style=unused] (15) to (14);
		\draw [style=unused] (15) to (11);
		\draw [style=unused] (0) to (8);
		\draw [style=unused] (1) to (9);
		\draw [style=unused] (9) to (12);
		\draw [style=unused] (8) to (12);
		\draw [style=unused] (9) to (13);
		\draw [style=unused] (13) to (10);
		\draw [style=unused] (13) to (8);
		\draw [style=unused] (8) to (14);
		\draw [style=unused] (14) to (10);
		\draw [style=unused] (14) to (9);
		\draw [style=unused] (8) to (11);
		\draw [style=unused] (9) to (11);
		\draw [style=unused] (11) to (10);
		\draw [style=unused] (6) to (1);
		\draw [style=simple] (3) to (2);
		\draw [style=simple] (1) to (4);
		\draw [style=simple] (0) to (4);
		\draw [style=simple] (4) to (2);
		\draw [style=simple] (5) to (2);
		\draw [style=simple] (5) to (0);
		\draw [style=simple] (6) to (2);
		\draw [style=simple] (0) to (3);
		\draw [style=simple] (1) to (3);
		\draw [style=simple] (7) to (4);
		\draw [style=simple] (7) to (5);
		\draw [style=simple] (7) to (6);
		\draw [style=simple] (7) to (3);
		\draw [style=simple] (15) to (12);
		\draw [style=embedding] (0) to (6);
		\draw [style=embedding] (7) to (15);
		\draw [style=embedding] (2) to (10);
		\draw [style=embedding] (1) to (5);
		\draw [style=embedding] (12) to (10);
	\end{pgfonlayer}
\end{tikzpicture}
}}
\par\end{center}
\vspace{-6mm}

\begin{center}
\subfloat[\emph{Double $K_{4}$}]{\resizebox{!}{\chimeraFigSize}{\begin{tikzpicture}
	\begin{pgfonlayer}{nodelayer}
		\node [style={emb_auxiliary_qubit}] (0) at (0, 2) {};
		\node [style={emb_logical_qubit}] (1) at (0, 1) {};
		\node [style={unused_qubit}] (2) at (0, -2) {};
		\node [style={emb_logical_qubit}] (3) at (1, 0) {};
		\node [style={emb_logical_qubit}] (4) at (-1, 0) {};
		\node [style={unused_qubit}] (5) at (-2, 0) {};
		\node [style={emb_logical_qubit}] (6) at (0, -1) {};
		\node [style={auxiliary_qubit}] (7) at (2, 0) {};
		\node [style={emb_auxiliary_qubit}] (8) at (5, 2) {};
		\node [style={emb_logical_qubit}] (9) at (5, 1) {};
		\node [style={unused_qubit}] (10) at (5, -2) {};
		\node [style={emb_logical_qubit}] (11) at (4, 0) {};
		\node [style={emb_logical_qubit}] (12) at (6, 0) {};
		\node [style={unused_qubit}] (13) at (7, 0) {};
		\node [style={emb_logical_qubit}] (14) at (5, -1) {};
		\node [style={auxiliary_qubit}] (15) at (3, 0) {};
	\end{pgfonlayer}
	\begin{pgfonlayer}{edgelayer}
		\draw [style=unused] (3) to (2);
		\draw [style=unused] (0) to (5);
		\draw [style=unused] (5) to (2);
		\draw [style=unused] (5) to (1);
		\draw [style=unused] (4) to (2);
		\draw [style=unused] (6) to (4);
		\draw [style=unused] (6) to (5);
		\draw [style=unused] (7) to (2);
		\draw [style=unused] (11) to (10);
		\draw [style=unused] (8) to (13);
		\draw [style=unused] (13) to (10);
		\draw [style=unused] (13) to (9);
		\draw [style=unused] (12) to (10);
		\draw [style=unused] (14) to (12);
		\draw [style=unused] (14) to (13);
		\draw [style=unused] (15) to (10);
		\draw [style=simple] (1) to (3);
		\draw [style=simple] (0) to (3);
		
		\draw [style=embedding] (1) to (4);
		
		\draw [style=simple] (4) to (0);
		
		\draw [style=embedding] (6) to (3);

		\draw [style=simple] (0) to (7);
		\draw [style=simple] (6) to (7);
		
		\draw [style=simple] (1) to (7);
		\draw [style=simple] (9) to (11);
		\draw [style=simple] (8) to (11);

		\draw [style=embedding] (9) to (12);
		
		\draw [style=simple] (12) to (8);
		
		\draw [style=embedding] (14) to (11);
		
		\draw [style=simple] (8) to (15);
		\draw [style=simple] (14) to (15);
		
		\draw [style=simple] (9) to (15);
		\draw [style={embedding_aux}] (0) to (8);
	\end{pgfonlayer}
\end{tikzpicture}
}}
\par\end{center}
\vfill{}
\end{minipage}~~~~~%
\begin{minipage}[t]{0.33\textwidth}%
\begin{center}
\subfloat[\emph{$K_{6}$}]{\rotatebox{270}{\resizebox{!}{0.93\textwidth}{\begin{tikzpicture}
	\begin{pgfonlayer}{nodelayer}
		\node [style={emb_auxiliary_qubit}] (0) at (2, 0) {};
		\node [style={emb_auxiliary_qubit}] (1) at (1, 0) {};
		\node [style={emb_logical_qubit}] (2) at (-2, 0) {};
		\node [style={emb_logical_qubit}] (3) at (0, -2) {};
		\node [style={emb_logical_qubit}] (4) at (0, -1) {};
		\node [style={emb_auxiliary_qubit}] (5) at (0, 1) {};
		\node [style={emb_auxiliary_qubit}] (6) at (0, 2) {};
		\node [style={emb_logical_qubit}] (7) at (-1, 0) {};
		\node [style={unused_qubit}] (8) at (2, 5) {};
		\node [style={unused_qubit}] (9) at (1, 5) {};
		\node [style={emb_logical_qubit}] (10) at (-2, 5) {};
		\node [style={unused_qubit}] (11) at (0, 7) {};
		\node [style={emb_logical_qubit}] (12) at (0, 6) {};
		\node [style={unused_qubit}] (13) at (0, 4) {};
		\node [style={unused_qubit}] (14) at (0, 3) {};
		\node [style={emb_logical_qubit}] (15) at (-1, 5) {};
		\node [style={unused_qubit}] (16) at (4, 0) {};
		\node [style={unused_qubit}] (17) at (5, 0) {};
		\node [style={emb_logical_qubit}] (18) at (8, 0) {};
		\node [style={emb_logical_qubit}] (19) at (6, -2) {};
		\node [style={emb_logical_qubit}] (20) at (6, -1) {};
		\node [style={unused_qubit}] (21) at (6, 1) {};
		\node [style={unused_qubit}] (22) at (6, 2) {};
		\node [style={unused_qubit}] (23) at (7, 0) {};
	\end{pgfonlayer}
	\begin{pgfonlayer}{edgelayer}
		\draw [style=unused] (17) to (21);
		\draw [style=unused] (21) to (18);
		\draw [style=unused] (21) to (16);
		\draw [style=unused] (16) to (22);
		\draw [style=unused] (22) to (18);
		\draw [style=unused] (22) to (17);
		\draw [style=unused] (16) to (19);
		\draw [style=unused] (17) to (19);
		\draw [style=unused] (23) to (20);
		\draw [style=unused] (23) to (21);
		\draw [style=unused] (23) to (22);
		\draw [style=unused] (23) to (19);
		\draw [style=unused] (5) to (0);
		\draw [style=unused] (11) to (10);
		\draw [style=unused] (9) to (12);
		\draw [style=unused] (8) to (12);
		\draw [style=unused] (9) to (13);
		\draw [style=unused] (13) to (10);
		\draw [style=unused] (13) to (8);
		\draw [style=unused] (8) to (14);
		\draw [style=unused] (14) to (10);
		\draw [style=unused] (14) to (9);
		\draw [style=unused, in=-44, out=135] (8) to (11);
		\draw [style=unused] (9) to (11);
		\draw [style=unused] (15) to (13);
		\draw [style=unused] (15) to (14);
		\draw [style=unused] (15) to (11);
		\draw [style=unused] (0) to (8);
		\draw [style=unused] (1) to (9);
		\draw [style=unused] (17) to (20);
		\draw [style=unused] (16) to (20);
		\draw [style=simple] (3) to (2);
		\draw [style=simple] (1) to (4);
		\draw [style=simple] (0) to (4);
		\draw [style=simple] (4) to (2);
		\draw [style={embedding_aux}] (1) to (5);
		\draw [style=simple] (5) to (2);
		\draw [style={embedding_aux}] (0) to (6);
		\draw [style=simple] (6) to (2);
		\draw [style=simple] (6) to (1);
		\draw [style=simple] (0) to (3);
		\draw [style=simple] (1) to (3);
		\draw [style=simple] (7) to (4);
		\draw [style=simple] (7) to (5);
		\draw [style=simple] (7) to (6);
		\draw [style=simple] (7) to (3);
		\draw [style=embedding] (12) to (10);
		\draw [style=simple] (15) to (12);
		\draw [style=embedding] (7) to (15);
		\draw [style=embedding] (2) to (10);
		\draw [style=embedding] (19) to (18);
		\draw [style=simple] (20) to (18);
		\draw [style=embedding] (3) to (19);
		\draw [style=embedding] (4) to (20);
	\end{pgfonlayer}
\end{tikzpicture}
}}

}
\par\end{center}%
\vfill{}
\end{minipage}\hfill{}
\end{figure}

\vspace{-2mm}

\subsection{Pegasus graph}

\vspace{-4mm}

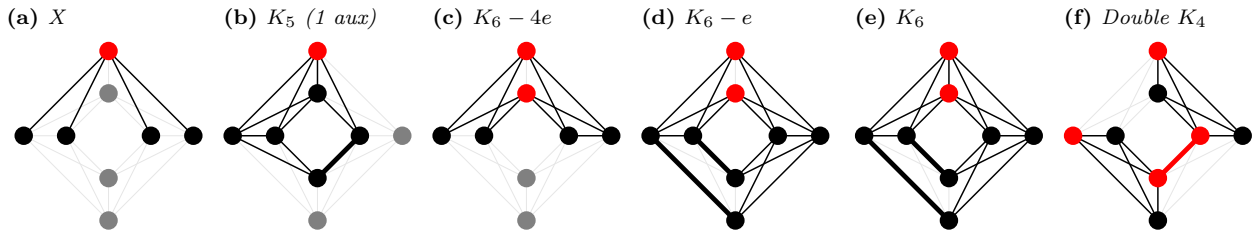
\begin{figure}[H]
\caption{{\small{}Minor embeddings of all }\textbf{\emph{\small{}\uline{quartic}}}{\small{}
to quadratic gadgets for single terms, onto a Pegasus `cell'. Grey
vertices and edges are not used.} }

\newcommand{\chimeraFigSize}{2.5cm}
\centering{}\subfloat[\emph{X}]{

\resizebox{!}{\chimeraFigSize}{\begin{tikzpicture}
	\begin{pgfonlayer}{nodelayer}
		\node [style=logical_qubit] (0) at (-2, -0) {};
		\node [style=logical_qubit] (1) at (-1, -0) {};
		\node [style=logical_qubit] (2) at (1, -0) {};
		\node [style=logical_qubit] (3) at (2, -0) {};
		\node [style=auxiliary_qubit] (4) at (0, 2) {};
		\node [style=unused_qubit] (5) at (0, 1) {};
		\node [style=unused_qubit] (6) at (0, -1) {};
		\node [style=unused_qubit] (7) at (0, -2) {};
	\end{pgfonlayer}
	\begin{pgfonlayer}{edgelayer}
		\draw [style=unused] (1) to (5);
		\draw [style=unused] (2) to (5);
		\draw [style=unused] (0) to (5);
		\draw [style=unused] (5) to (3);
		\draw [style=unused] (1) to (6);
		\draw [style=unused] (6) to (2);
		\draw [style=unused] (6) to (3);
		\draw [style=unused] (6) to (0);
		\draw [style=unused] (0) to (7);
		\draw [style=unused] (7) to (3);
		\draw [style=unused] (7) to (2);
		\draw [style=unused] (7) to (1);
		\draw [style=simple] (0) to (4);
		\draw [style=simple] (1) to (4);
		\draw [style=simple] (2) to (4);
		\draw [style=simple] (4) to (3);
		
		\draw [style={unused_added}] (0) to (1);
		\draw [style={unused_added}] (2) to (3);

		\draw [style={unused_added}] (4) to (5);
		\draw [style={unused_added}] (6) to (7);
	\end{pgfonlayer}
\end{tikzpicture}}}~~\subfloat[$K_{5}$ \emph{(1 aux)}]{\resizebox{!}{\chimeraFigSize}{\begin{tikzpicture}
	\begin{pgfonlayer}{nodelayer}
		\node [style=logical_qubit] (0) at (-2, -0) {};
		\node [style=logical_qubit] (1) at (-1, -0) {};
		\node [style=unused_qubit] (2) at (2, -0) {};
		\node [style=auxiliary_qubit] (3) at (0, 2) {};
		\node [style=logical_qubit] (4) at (0, 1) {};
		\node [style={emb_logical_qubit}] (5) at (0, -1) {};
		\node [style=unused_qubit] (6) at (0, -2) {};
		\node [style={emb_logical_qubit}] (7) at (1, -0) {};
	\end{pgfonlayer}
	\begin{pgfonlayer}{edgelayer}
		\draw [style=unused] (3) to (2);
		\draw [style=unused] (4) to (2);
		\draw [style=unused] (0) to (6);
		\draw [style=unused] (6) to (2);
		\draw [style=unused] (6) to (1);
		\draw [style=unused] (5) to (2);
		\draw [style=unused] (7) to (6);
		\draw [style=simple] (1) to (4);
		\draw [style=simple] (0) to (4);
		
		\draw [style=added] (0) to (1);
		\draw [style=simple] (1) to (5);
		
		\draw [style=simple] (5) to (0);
		
		\draw [style=added] (3) to (4);
		\draw [style={unused_added}] (5) to (6);
		\draw [style=simple] (0) to (3);
		\draw [style=simple] (1) to (3);
		\draw [style=simple] (7) to (4);
		\draw [style={embedding}] (7) to (5);
		
		\draw [style=simple] (7) to (3);
		\draw [style={unused_added}] (7) to (2);
	\end{pgfonlayer}
\end{tikzpicture}}}~~\subfloat[\emph{$K_{6}-4e$}]{\resizebox{!}{\chimeraFigSize}{\begin{tikzpicture}
	\begin{pgfonlayer}{nodelayer}
		\node [style=logical_qubit] (0) at (-2, -0) {};
		\node [style=logical_qubit] (1) at (-1, -0) {};
		\node [style=logical_qubit] (2) at (2, -0) {};
		\node [style=auxiliary_qubit] (3) at (0, 2) {};
		\node [style=auxiliary_qubit] (4) at (0, 1) {};
		\node [style=unused_qubit] (5) at (0, -1) {};
		\node [style=unused_qubit] (6) at (0, -2) {};
		\node [style=logical_qubit] (7) at (1, -0) {};
	\end{pgfonlayer}
	\begin{pgfonlayer}{edgelayer}
		\draw [style=simple] (3) to (2);
		\draw [style=simple] (1) to (4);
		\draw [style=simple] (0) to (4);
		\draw [style=simple] (4) to (2);
		\draw [style={unused_added}] (0) to (1);
		\draw [style=unused] (1) to (5);
		\draw [style=unused] (5) to (2);
		\draw [style=unused] (5) to (0);
		\draw [style=unused] (0) to (6);
		\draw [style=unused] (6) to (2);
		\draw [style=unused] (6) to (1);
		\draw [style={unused_added}] (3) to (4);
		\draw [style={unused_added}] (5) to (6);
		\draw [style=simple] (0) to (3);
		\draw [style=simple] (1) to (3);
		\draw [style=simple] (7) to (4);
		\draw [style=unused] (7) to (5);
		\draw [style=unused] (7) to (6);
		\draw [style=simple] (7) to (3);
		\draw [style=added] (7) to (2);
	\end{pgfonlayer}
\end{tikzpicture}}}~~\subfloat[\emph{$K_{6}-e$}]{\resizebox{!}{\chimeraFigSize}{\begin{tikzpicture}
	\begin{pgfonlayer}{nodelayer}
		\node [style=emb_logical_qubit] (0) at (-2, -0) {};
		\node [style=emb_logical_qubit] (1) at (-1, -0) {};
		\node [style=logical_qubit] (2) at (2, -0) {};
		\node [style=auxiliary_qubit] (3) at (0, 2) {};
		\node [style=auxiliary_qubit] (4) at (0, 1) {};
		\node [style=emb_logical_qubit] (5) at (0, -1) {};
		\node [style=emb_logical_qubit] (6) at (0, -2) {};
		\node [style=logical_qubit] (7) at (1, -0) {};
	\end{pgfonlayer}
	\begin{pgfonlayer}{edgelayer}
		\draw [style=simple] (3) to (2);
		\draw [style=simple] (1) to (4);
		\draw [style=simple] (0) to (4);
		\draw [style=simple] (4) to (2);
		\draw [style=added] (0) to (1);
		\draw [style={embedding}] (1) to (5);
		\draw [style=simple] (5) to (2);
		\draw [style=unused] (5) to (0);
		\draw [style={embedding}] (0) to (6);
		\draw [style=simple] (6) to (2);
		\draw [style=unused] (6) to (1);
		\draw [style={unused_added}] (3) to (4);
		\draw [style={unused_added}] (5) to (6);
		\draw [style=simple] (0) to (3);
		\draw [style=simple] (1) to (3);
		\draw [style=simple] (7) to (4);
		\draw [style=simple] (7) to (5);
		\draw [style=simple] (7) to (6);
		\draw [style=simple] (7) to (3);
		\draw [style=added] (7) to (2);
	\end{pgfonlayer}
\end{tikzpicture}
}}~~\subfloat[\emph{$K_{6}$}]{{\tiny{}}\resizebox{!}{\chimeraFigSize}{\begin{tikzpicture}
	\begin{pgfonlayer}{nodelayer}
		\node [style=emb_logical_qubit] (0) at (-2, -0) {};
		\node [style=emb_logical_qubit] (1) at (-1, -0) {};
		\node [style=logical_qubit] (2) at (2, -0) {};
		\node [style=auxiliary_qubit] (3) at (0, 2) {};
		\node [style=auxiliary_qubit] (4) at (0, 1) {};
		\node [style=emb_logical_qubit] (5) at (0, -1) {};
		\node [style=emb_logical_qubit] (6) at (0, -2) {};
		\node [style=logical_qubit] (7) at (1, -0) {};
	\end{pgfonlayer}
	\begin{pgfonlayer}{edgelayer}
		\draw [style=simple] (3) to (2);
		\draw [style=simple] (1) to (4);
		\draw [style=simple] (0) to (4);
		\draw [style=simple] (4) to (2);
		\draw [style=added] (0) to (1);
		\draw [style={embedding}] (1) to (5);
		\draw [style=simple] (5) to (2);
		\draw [style=unused] (5) to (0);
		\draw [style={embedding}] (0) to (6);
		\draw [style=simple] (6) to (2);
		\draw [style=unused] (6) to (1);
		\draw [style={unused_added}] (5) to (6);
		\draw [style=simple] (0) to (3);
		\draw [style=simple] (1) to (3);
		\draw [style=simple] (7) to (4);
		\draw [style=simple] (7) to (5);
		\draw [style=simple] (7) to (6);
		\draw [style=simple] (7) to (3);
		\draw [style=added] (7) to (2);
		\draw [style=added] (3) to (4);
	\end{pgfonlayer}
\end{tikzpicture}}}~~\subfloat[\emph{Double $K_{4}$}]{\resizebox{!}{\chimeraFigSize}{\begin{tikzpicture}
	\begin{pgfonlayer}{nodelayer}
		\node [style={logical_qubit}] (0) at (2, 0) {};
		\node [style={emb_auxiliary_qubit}] (1) at (1, 0) {};
		\node [style={auxiliary_qubit}] (2) at (-2, 0) {};
		\node [style={logical_qubit}] (3) at (0, -2) {};
		\node [style={emb_auxiliary_qubit}] (4) at (0, -1) {};
		\node [style={logical_qubit}] (5) at (0, 1) {};
		\node [style={auxiliary_qubit}] (6) at (0, 2) {};
		\node [style={logical_qubit}] (7) at (-1, 0) {};
	\end{pgfonlayer}
	\begin{pgfonlayer}{edgelayer}
		\draw [style=simple] (3) to (2);
		\draw [style={embedding_aux}] (1) to (4);
		\draw [style=unused] (0) to (4);
		\draw [style=simple] (4) to (2);
		\draw [style=added] (0) to (1);
		\draw [style=simple] (1) to (5);
		\draw [style=unused] (5) to (2);
		\draw [style=simple] (5) to (0);
		\draw [style=simple] (0) to (6);
		\draw [style=unused] (6) to (2);
		\draw [style=simple] (6) to (1);
		\draw [style=added] (5) to (6);
		\draw [style=unused] (0) to (3);
		\draw [style=unused] (1) to (3);
		\draw [style=simple] (7) to (4);
		\draw [style=unused] (7) to (5);
		\draw [style=unused] (7) to (6);
		\draw [style=simple] (7) to (3);
		\draw [style=added] (7) to (2);
		\draw [style=added] (3) to (4);
	\end{pgfonlayer}
\end{tikzpicture}}}
\end{figure}

\section{Recommended gadgets}

When reducing the number of total qubits is the most important factor
in compiling a discrete optimization problem, negative terms (whether
cubic or quartic) can be quadratized with only 1 auxiliary qubit,
and in such a way not to require any additional qubits for embedding
onto either Chimera and Pegasus (so only 1 auxiliary qubit is required
in total). 

For positive terms the situation is a bit more complicated: Still
only 1 auxiliary qubit is required for quadratization of cubic terms,
and for Pegasus the quadratization can be done in such a way as to
require no further auxiliary qubits for the minor-embedding, however
embedding with Chimera will require at minimum 1 further auxiliary
qubit (so 1 auxiliary qubit is required in total for Pegasus, but
2 auxiliary qubits are required in total for Chimera). 

For quartic positive terms, we need at least 2 total auxiliary qubits
on Pegasus and at least 3 auxiliary qubits for Chimera. It is interesting
to note that for positive quartic terms, while the gadgets leading
to the $K_{5}$ graph require 1 fewer auxiliary qubit than the gadgets
leading to the $K_{6}-4e$ graph for quadratization, the graph needs
2 more auxiliary qubits than $K_{6}-4e$ for embedding onto Chimera,
and 1 more auxiliary qubit for embedding onto Pegasus (which requires
0 in the former case). Therefore, while gadgets leading to $K_{5}$
may appear at first sight to be more efficient than gadgets leading
to $K_{6}-4e$ due to requiring fewer qubits for quadratization, they
require more \emph{total }auxiliary qubits for embedding onto Chimera
and an equivalent number of total qubits for embedding onto Pegasus.

\section*{Acknowledgments}

We gratefully thank Nick Wormald of Monash University for helpful
comments about our gadget graphs. We gratefully thank Szilard Szalay
from the Wigner Research Centre for Physics of the Hungarian Academy
of Sciences for assistance with making the graphs in Figures 1 and
2. NC was funded by EPSRC (Project: EP/S00114X).

\selectlanguage{british}%
\bibliographystyle{apsrev4-1}
\bibliography{\string"/home/nike/pCloud Sync/library\string"}
\selectlanguage{english}%

\end{document}